\documentclass[%
 reprint,
 amsmath,amssymb,
 aps,
]{revtex4-1}

\usepackage{graphicx}
\usepackage{dcolumn}
\usepackage{bm}
\usepackage{dsfont}
\usepackage{appendix}

\usepackage{color}
\usepackage{ulem}

\begin{document}

\preprint{APS/123-QED}

\title{Note on the Witten index and the dynamical breaking of a supersymmetric gauge theory }

\author{Renata Jora
	$^{\it \bf a}$~\footnote[1]{Email:
		rjora@theory.nipne.ro}}

\affiliation{$^{\bf \it a}$ National Institute of Physics and Nuclear Engineering PO Box MG-6, Bucharest-Magurele, Romania}

\email[E-mail: ]{rjora@theory.nipne.ro}

\begin{abstract}

It is known that in order for a supersymmetric gauge theory to be dynamically broken a necessary but not sufficient condition is that the Witten index $(-1)^F=n_b^0-n_f^0$ is zero. In the case the Witten index is zero it is in general hard to distinguish between theories with or without dynamical supersymmetry breaking. In this work we show that  for a large class of theories the cancellation of the Witten index by itself signals the formation of a quark condensate and therefore produces dynamical supersymmetry breaking.

\end{abstract}
\maketitle

With the advent of the LHC experiments \cite{Atlas}, \cite{CMS} the particle physics has entered an era of unprecedented experimental confirmation and search of beyond of the standard model particle and interactions. Despite the absence of any hint of supersymmetry in the LHC data supersymmetric theories remain largely the most important candidates for extensions of the UV sector of the standard model and for solving an array of problems in particle physics and not only. If supersymmetry exists it must be broken. It would be then of utmost interest to find or construct a theory is dynamically broken through non-perturbative effects.

According to Witten constraint \cite{Witten1}, \cite{Witten2} a necessary but not sufficient condition for dynamical supersymmetry breaking is given by:
\begin{eqnarray}
(-1)^F=n_b^0-n_F^0=0,
\label{wittenconstr4555}
\end{eqnarray}
where $(-1)^F$ is the Witten index, $n_B^0$ is the number of bosonic modes and $n_F^0$ is the number of fermionic modes. Because of the supersymmetry all non zero modes are paired up between bosons and fermions and they cancel up.

The Witten index can be calculated easily for supersymmetric $SU(N)$ supersymmetric Yang Mills theory \cite{Witten1} and it is $(-1)^F=N$ and for the theories with massive matter (because in the end the matter can be integrated out) but it is ill-defined and hard to calculate for massless theories with flat directions. The latter correspond to those $D$-directions of non zero $\Phi$, the squark for which the squark potential vanishes. In particular if the Witten index is zero the theory displays dynamical supersymmetry breaking if $n_b^0=n_F^0=0$ and remains supersymmetrical if $n_B^0=n_F^0\neq0$.

The nonperturbative dynamics of many supersymmetric gauge theories with or without flat directions has been elucidated in a series of groundbreaking works \cite{Seiberg1}, \cite{Seiberg2} (and the references therein).

This work is written with the aim of filling a gap of knowledge regarding the instances when the Witten index is zero. We expect that a theory with the Witten index zero must contain matter in an arbitrary representation of the gauge group with or without an additional flavor symmetry.  Whereas in the presence of a chiral symmetry this might be broken by the vacuum expectation values of the squarks without breaking supersymmetry the presence of the quark condensate always indicates the dynamical breaking of the supersymmetry \cite{Witten2}. In this paper we will show under general and mild assumptions that a zero Witten index in case this is well defined always signals the formation of a quark condensate $\bar{\Psi}\Psi$ and thus  dynamical supersymmetry breaking.

We consider a supersymmetric gauge theory with matter in an arbitrary representation of the gauge group. We are interested to see in what circumstances a quark vacuum condensate may form. For simplicity we will assume that a Dirac mass term may be introduced in the theory without the associated squarks mass term. A small mass term only for quarks breaks explicitly the supersymmetry.  We will further assume reasonably that the limit $m\rightarrow 0$ makes sense.

We denote the Hamiltonian of the supersymmetric theory with the mass term set to zero $H_0$. Then,
\begin{eqnarray}
H=H_0+ \lambda H_1,
\label{newhamilt75664}
\end{eqnarray}
where $\lambda$ is a small parameter and $H_1$ for example is of the type $ \int d^3x m\bar{\Psi}\Psi$ (with $m$ small but finite) or any other bilinear that may receive mass depending on the theory. Our arguments may extend to any theory that might contain sensible mass terms for the quarks even if this term breaks or not some symmetry.

The partition function for the supersymmetric theory with the hamiltonian $H$ is:
\begin{eqnarray}
Z=\int dX \exp[i\int d^4 x{\cal L}]=
\frac{\det \langle F |H|F\rangle}{\det \langle B|H|B\rangle}.
\label{hamilt75665}
\end{eqnarray}
Here the path integral is taken over all degrees of freedom named generically $X$, $F$ are the fermionic degrees of freedom and $B$ are the bosonic degrees of freedom.

Then the quantity:
\begin{eqnarray}
\frac {d \ln Z}{d \lambda }|_{\lambda=0}=-i\int d^4 x\langle m\bar{\Psi}\Psi\rangle=c,
\label{cond664554}
\end{eqnarray}
makes sense in the quantum approach.  We shall consider that if $\langle \bar{\Psi}\Psi\rangle\neq 0$ the integral is regularized to yield a finite answer and if $\langle \bar{\Psi}\Psi\rangle=0$ the integral is zero.

We assume that the Witten index is zero. Then the number of fermionic degrees of freedom, including the zero modes is equal to the number of bosonic degrees of freedom and one may write:
\begin{eqnarray}
&&c=\frac{d\ln Z}{d \lambda}|_{\lambda=0}={\rm Tr}\Bigg[\frac{\langle F|H_1|F\rangle}{\langle F|H_0|F\rangle}-\frac{\langle B|H_1|B\rangle}{\langle B|H_0| B\rangle}\Bigg]=
\nonumber\\
&&={\rm Tr}\Bigg[\frac{\langle F|H'|F\rangle}{\langle F|H_0|F\rangle}-\frac{\langle B|H'|B\rangle}{\langle B|H_0| B\rangle}\Bigg],
\label{res66455}
\end{eqnarray}
where $H'=H_0+H_1$.
Here we used the fact that all eigenvalues for the hamiltonian $H_0$, including those equal to zero are paired between the bosonic and fermionic states and that,
\begin{eqnarray}
{\rm Tr}\Bigg[\frac{\langle F|H_0|F\rangle}{\langle F|H_0|F\rangle}-\frac{\langle B|H_0|B\rangle}{\langle B|H_0| B\rangle}\Bigg]=0.
\label{ani847465}
\end{eqnarray}

 All vacuum expectation operators  are taken on the eigenmodes of the unperturbed hamiltonian $H_0$. Eq. (\ref{res66455}) may be rewritten as:
\begin{eqnarray}
c=-\sum_n\Bigg[\frac{E_n'}{E_n}\Bigg],
\label{eigenv8857746}
\end{eqnarray}
where $E_n$ are the eigenvalues of the hamiltonian $H_0$ and $E_n'$ are the eigenvalues of the hamiltonian $H'$. We considered non-degenerate perturbation theory with a perturbation given by the small mass $m$  but the same arguments may apply with some minor difficulties to the degenerate one.

But the perturbed hamiltonian has the expression,
\begin{eqnarray}
H'=H_0+\int d^3 xm \bar{\Psi}(x)\Psi(x),
\label{perthamilt7586776}
\end{eqnarray}
with,
\begin{eqnarray}
&&H_1=\int d^3 xm \bar{\Psi}(x)\Psi(x)=
\nonumber\\
&&\int \frac{d^3 q}{(2\pi)^3}\frac{m^2}{E_p}[\sum_sa^{s\dagger}_{{\bf q}}a^s_{\bf q}+b^{s\dagger}_{{\bf q}}b^s_{{\bf q}}]-
\nonumber\\
&&2\frac{m^2}{E_p}V\int\frac{d^3q}{(2\pi)^3},
\label{hamiltexpr6564}
\end{eqnarray}
which is defined up to a negative infinite constant (Here $a^s_{{\bf q}}$, $b^s_{{\bf q}}$ are the standard annihilation fermion operators whereas $a^{s\dagger}_{{\bf q}}$, $b^{s\dagger}_{{\bf q}}$ are the corresponding creation operators). The infinite constant can either be disregarded or a constant term that cancel it can be introduced in the Lagrangian from the beginning. This should not affect in any way the quantum behavior of the theory. Thus it is clear that the contribution of the perturbation on the states of hamiltonian $H_0$ is positive or zero, therefore
the eigenvalues of $H'$ are positive or zero.

Since the Hamiltonian is semi positive definite  the sum in Eq. (\ref{eigenv8857746}) cannot be zero and since $c$ should be regularized neither infinite. Thus also no $E_n$ can be zero and $c\neq 0$.

If the Witten index is not zero then $c$ remains undetermined  and one cannot say anything about its value in this approach.

In this work we showed that if the Witten index of a supersymmetric gauge theory that can be calculated is exactly zero than the theory contains at least one quark condensate and the supersymmetry is broken by non-perturbative effects.

It is known that in all instances when the supersymmetry is dynamically broken the Witten index is zero but the reversed is not necessarily true. Here we proved that for a large class of theories the Witten constraint is not only necessary but also sufficient for the theory to be dynamically broken.  Our method applies to such supersymmetric gauge theories that contain matter that may lead to mass terms for fermion bilinears that are Lorentz scalars and that may be or not   gauge singlets. For example for $SU(N)$ with fermions in the fundamental or antifundamental representation it may apply to the regime where $N_f<N$ where arguably supersymmetry is dynamically broken. It can also be applied  to $SU(5)$ with matter in the antisymmetric ($10$) and antifundamental ($5^*$) representations or $SO(10)$ with matter in the spinor representation ($16$).  It was shown \cite{Murayama} that indeed the associated low energy effective theories   display dynamical supersymmetry breaking. In all these cases studied through an effective superpotential approach in the low energy regime the gauge symmetry is broken to a smaller subgroup fact which  is perfectly compatible  with the formation of fermion condensates. However in order for our method to be verified one needs a prior  knowledge of the Witten index.

But for many theories the calculation of the Witten index must be associated to some knowledge of the underlying non-perturbative dynamics of the gauge theory.  The Witten constraint for supersymmetry along with the result obtained in this work should thus be taken in conjunction with other non-perturbative methods like holomorphicity, duality or the weakly interacting limit to lead to fruitful prediction of the dynamical supersymmetry breaking. Our work is a step further in the as yet not completely chartered territory of the non-perturbative behavior of a supersymmetric gauge theory.

\end{document}